\documentstyle[preprint,aps,epsfig]{revtex}
\tightenlines
\begin{document}
\draft

\title{Strong nuclear enhancement in intermediate mass Drell-Yan
       production}
\author{Jianwei Qiu$^a$ and Xiaofei Zhang$^{a,b}$}
\address{
 $^a$Department of Physics and Astronomy,
     Iowa State University \\
     Ames, Iowa 50011, USA \\
 $^b$Physics Department,
     Kent State University \\
     Kent, Ohio 44242, USA
}

\date{September 21, 2001}
\maketitle
\begin{abstract}
We calculate nuclear effect in Drell-Yan massive lepton-pair
production in terms of parton multiple scattering in Quantum
Chromodynamics (QCD).  We present the nuclear modification to
inclusive Drell-Yan cross section $d\sigma/dQ^2$ in terms of
multiparton correlation functions.  By extracting the size of the
correlation functions from measured Drell-Yan transverse momentum
broadening in nuclear medium, we determine the nuclear modification at
$O(\alpha_s/Q^2)$.  We find that the nuclear modification strongly
enhances the inclusive Drell-Yan cross section in the intermediate
mass region (IMR): $1.5 < Q < 2.5$~GeV.  We argue that the Drell-Yan
process is responsible for the majority of the observed dimuon
enhancement in the IMR in the Pb-Pb collisions at CERN SPS. 
\end{abstract}
\vspace{0.2in}



The inclusive Drell-Yan massive lepton-pair production in
hadron-nucleus and nucleus-nucleus collisions is an excellent
laboratory for theoretical and experimental investigations of the role
of nuclear medium in strong interaction dynamics.  The massive
lepton-pair production is also a channel for studying
nuclear medium effect in production of quarkonium states and
intermediate vector bosons.  Recent measurements of massive
lepton-pair production in both hadron-nucleus and nucleus-nucleus
collisions show a very little nuclear dependence in Drell-Yan
continuum, $d\sigma/dQ$, for
lepton pair mass $Q$ larger than 4~GeV \cite{E772-dy,CERN-dy}.
Because of this lack of nuclear dependence in
Drell-Yan continuum, the reported strong J/$\psi$ suppression
by the NA50 Collaboration was normalized to the Drell-Yan continuum in
the lepton pair mass range $2.9 < Q < 4.5$~GeV \cite{NA50-jpsi}.

Recently, it was shown by the NA38 and NA50 Collaborations that muon
pair production for dimuon invariant mass between the $\phi$ and the
J/$\psi$ in heavy nucleus-nucleus collisions exceeds the expectation
based on a linear extrapolation of the $p-A$ sources with the product
of the mass numbers of the projectile and target nuclei
\cite{NA38NA50-dimuon}.  The excess increases with the number of
participant nucleons, and the ratio between the observed dimuon yield
and the expected sources reaches a factor of 2 for central Pb-Pb
collisions.  There have been a lot of effort to attribute such an
excess to the sources other than the Drell-Yan continuum, such as the
enhancement of open charm production \cite{OpenCharm}, thermal
dimuons production \cite{thermal-l}, and secondary
meson-meson scattering in nuclear medium \cite{Meson-meson}.  As shown
in Ref.~\cite{NA38NA50-dimuon}, the Drell-Yan continuum is the
dominant source of the dilepton production in this relevant mass
range in Pb-Pb collisions with 158~GeV/nucleon beam.  An enhancement
in the Drell-Yan continuum is much more effective than all other
sources for interpreting the observed excess. 
Since the J/$\psi$ cross section in Pb-Pb collisions was normalized to
the measured Drell-Yan continuum in the same experiment, it is
important to understand the nuclear dependence of the Drell-Yan
continuum in the same mass range \cite{NA50-jpsi}.
It is the purpose of this Letter to calculate the
nuclear modification to the Drell-Yan continuum in the mass range
between the $\phi$ and the J/$\psi$, which is reorganized as the
intermediate mass region (IMR).

Since we are interested in the Drell-Yan continuum, or in general,
Drell-Yan cross section at low
invariant mass, $Q$, we ignore contributions from the
intermediate vector boson $Z$.  Therefore, the inclusive Drell-Yan
cross section in a collision between hadrons,
$h(p)+h'(p')\rightarrow \gamma^*(\rightarrow l\bar{l}(Q))+X$, can be
expressed in terms of the cross section for production of an
unpolarized virtual photon of the same invariant mass
\cite{LamTung-dy,QZ-photon}
\begin{equation}
\frac{d\sigma}{d^4Q}
=\frac{1}{12 \pi^3}\, \frac{\alpha^2_{em}}{S^2Q^2} \,
W_{\mu\nu}(Q)\,P^{\mu\nu}(Q) \, ,
\label{dy-d4q}
\end{equation}
where $S=(p+p')^2$ is the c.m. energy of the collision.  The
$W_{\mu\nu}(Q)$ in Eq.~(\ref{dy-d4q}) is a hadronic tensor that
satisfies electromagnetic (EM) current conservation, $Q_\mu
W^{\mu\nu}(Q) = W^{\mu\nu}(Q) Q_\nu = 0$.  Because of the EM current
conservation of $W_{\mu\nu}(Q)$, the polarization tensor for the
virtual photon, $P^{\mu\nu}(Q) = \sum_{\lambda=T,L}\,
  \epsilon^{*\mu}_\lambda(Q)\, \epsilon^{\nu}_\lambda(Q)$
in Eq.~(\ref{dy-d4q}), is often replaced by $-g^{\mu\nu}$.  
The $T$ and $L$ here represent the 
transverse and longitudinal polarization of the virtual photon,
respectively.

\begin{figure}
\begin{center}
\epsfig{figure=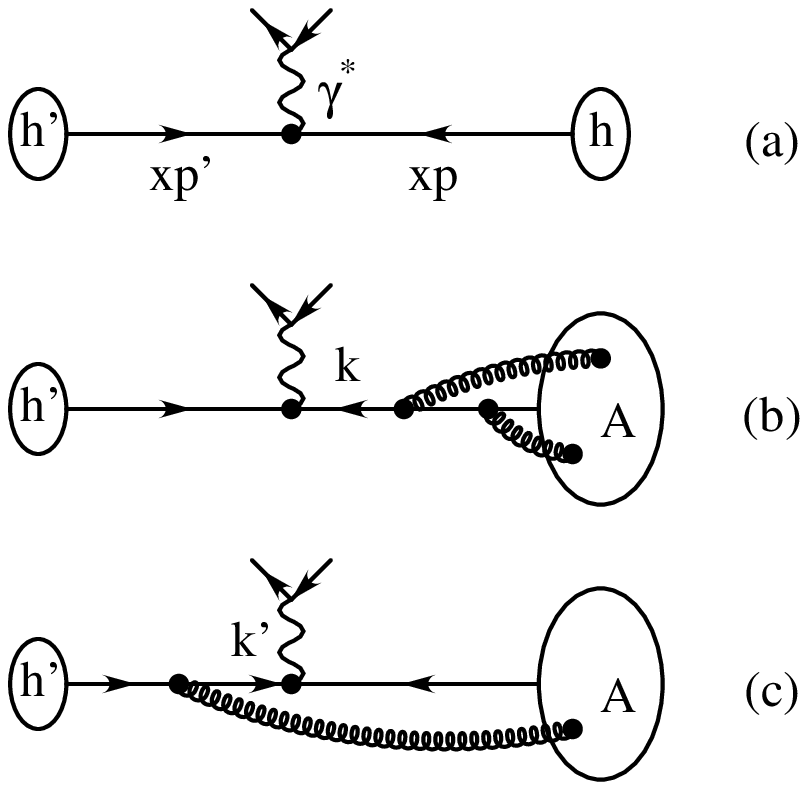,width=2.5in}
\end{center}
\caption{Sketch for Drell-Yan process in hadron-hadron and
hadron-nucleus collisions: (a) single hard scattering, (b) multiple
interaction internal to the nucleus, (c) multiple interaction affects
the hard scattering.}
\label{fig1}
\end{figure}

When energy exchange of the collision, $Q$, is much larger than the
typical momentum scales of initial-state hadron wave functions, the
Drell-Yan cross section, $d\sigma/d^4Q$, is dominated by a {\it
single} hard collision between a quark and an antiquark from two
incoming hadrons, as sketched in Fig.~\ref{fig1}(a).
Corrections due to interactions involving extra partons from
incoming hadrons are power suppressed by the ratio of momentum scale
of the hadron wave functions and the energy exchange of the hard
collision.  Therefore, leading power contribution to the Drell-Yan
cross section can be factorized as \cite{CSS-fac}
\begin{equation}
\frac{d\sigma_{hh'}}{d^4Q}
\approx 
\frac{d\sigma_{hh'}^{(s)}}{d^4Q}
= 
\sum_{a,b} \int dx\, \phi_{a/h}(x)
           \int dx'\, \phi_{b/h'}(x') \,
           \frac{d\hat{\sigma}_{ab}}{d^4Q}\, ,
\label{dy-d4q-fac}
\end{equation}
where superscript $(s)$ indicates a single hard scattering and
$\sum_{a,b}$ run over all parton flavors.  The $\phi$'s are
parton distributions and the factorization scale dependence are
suppressed.  The $d\hat{\sigma}_{ab}/d^4Q$ represents the
perturbatively calculable rate of partonic collision.  At the lowest
order of quark-antiquark annihilation, the inclusive Drell-Yan cross
section is given by \cite{DrellYan}
\begin{equation}
\frac{d\sigma_{hh'}^{(s)}}{dQ^2}
\approx 
  \left( \frac{4\pi \alpha_{em}^2}{9 S Q^2} \right)
  \sum_{q}\, e_q^2\,
  \int dy\, \phi_{q/h}(x)\, \phi_{\bar{q}/h'}(x')
\label{dy-dq2-s-lo}
\end{equation}
where the parton momentum fractions are given by
\begin{equation}
x=\frac{Q}{\sqrt{S}}\, {\rm e}^{+y}, 
\quad \mbox{and} \quad
x'=\frac{Q}{\sqrt{S}}\, {\rm e}^{-y}, 
\label{x-xp}
\end{equation}
with rapidity $y$.
The high order corrections in powers of $\alpha_s$ to the inclusive
cross section in Eq.~(\ref{dy-dq2-s-lo}) are well-defined in QCD
perturbation theory \cite{CTEQ-handbook}.  For hadron-hadron
collisions, Drell-Yan cross sections calculated in perturbative QCD
are consistent with all existing data for $Q>4$~GeV \cite{CTEQ5-pdf}.

In hadron-nucleus and nucleus-nucleus collisions, more partons are
available at a given impact parameter.  Consider the Drell-Yan process
in hadron-nucleus collisions.  Before the hard collision of producing
the lepton-pair, partons from different nucleons can either interact
between themselves, as sketched in Fig.~\ref{fig1}(b), or interact
with the incoming parton, as sketched in Fig.~\ref{fig1}(c).  After
taking a square of the scattering amplitude in Fig.~\ref{fig1}(b), the
leading contribution to the Drell-Yan cross section comes from the
phase space where the parton momentum $k$ is pinched, which
corresponds to the situation in which the interactions between partons
from different nucleons took place ``long'' before the short-distance
hard collision.  That is, only a single parton from the nucleus
participates the short-distance hard collision.  The parton-parton
interactions between nucleons are internal to the nucleus, and should
not affect the dynamics of short-distance hard collision.
Therefore, leading contributions from the multiple scattering process
in Fig.~\ref{fig1}(b) to the Drell-Yan cross section should have the
same factorized form as that in Eq.~(\ref{dy-d4q-fac}), except the
parton distributions of a free hadron are replaced by the parton
distributions of a nucleus.  Those interactions internal to the
nucleus in Fig.~\ref{fig1}(b) are responsible for the phenomena of
nuclear shadowing and EMC effect, which represent the differences
between parton distributions of a free nucleon and those inside a
nucleus.  The nuclear shadowing and EMC effect provide small nuclear
size dependence to the inclusive Drell-Yan cross section, or Drell-Yan
continuum, $d\sigma/dQ$. 

In this Letter, we concentrate on the contributions due to the type of
parton multiple scattering shown in Fig.~\ref{fig1}(c).  Unlike the
parton momentum $k$ in Fig.~\ref{fig1}(b), the parton momentum $k'$ in
Fig.~\ref{fig1}(c) is not pinched after we take a square of the
scattering amplitude \cite{QS-fac}.  Therefore, the type of double
scattering in Fig.~\ref{fig1}(c) is a part of the hard collision for
producing the lepton-pair.  Because of the extra initial-state
interaction, physical contributions from the double scattering are
suppressed by a power of $Q^2$, known as the high twist contributions
\cite{QS-dy-t4}.  Although they are normally small in hadron-hadron
collisions, such power suppressed contributions can be important in
hadron-nucleus and nucleus-nucleus collisions, because of enhancement
of nuclear size \cite{QS-hardprobe}.

Before presenting our calculation of the nuclear modification to the
Drell-Yan continuum caused by the type of multiple scattering in
Fig.~\ref{fig1}(c), we estimate the significance of such power
suppressed modification.  Let the inclusive Drell-Yan cross section in
the collisions between nucleus $A$ and nucleus $B$ be approximated as
\begin{eqnarray}
\frac{d\sigma_{AB}}{dQ^2} 
&\approx &
AB\frac{d\sigma_{NN}^{(s)}}{dQ^2}
+ \frac{d\sigma_{AB}^{(d)}}{dQ^2}
\nonumber \\
& \equiv &
AB\frac{d\sigma_{NN}^{(s)}}{dQ^2}
\left[ 1 + R_{AB}(Q) \right]\, ,
\label{dy-s-d}
\end{eqnarray}
where superscript $(d)$ represents the double scattering -- a source
of leading power corrections.  The
$d\sigma_{NN}^{(s)}/dQ^2$ is the single scattering cross section given
in Eq.~(\ref{dy-dq2-s-lo}) with parton distributions of free hadron
replaced by the effective parton distributions of a nucleon inside a
nucleus.  The $R_{AB}(Q)$ represents a ratio of the double scattering
over the single scattering, and has the following dependence
\cite{QS-hardprobe}
\begin{equation}
R_{AB}(Q) = \frac{d\sigma_{AB}^{(d)}/dQ^2}
                 {AB d\sigma_{NN}^{(s)}/dQ^2}
\propto \frac{\alpha_s(Q)}{Q^2} \langle L_{AB} \rangle
\label{rat-d-s}
\end{equation}
where $\langle L_{AB} \rangle \propto A^{1/3}+B^{1/3}$ is an effective
medium length.  Since $d\sigma/dQ = 2Q d\sigma/dQ^2$, the $R_{AB}(Q)$
in Eq.~(\ref{rat-d-s}) is also equal to the ratio of the Drell-Yan
continuum.  If we assume that the $R_{AB}(Q)$ is positive
and is about 5\% at $Q=4$~GeV for the Drell-Yan
continuum in hadron-nucleus collision, which is within the
experimental uncertainty of dilepton yield at that energy
\cite{NA38NA50-dimuon}, the ratio $R_{AB}(Q)$ will yield almost 50\%
enhancement to the Drell-Yan continuum in the central region of Pb-Pb
collisions at $Q=2$~GeV.  Such an enhancement could explain the
majority of the dimuon excess observed in Pb-Pb collisions.

In order to quantitatively understand the strong nuclear enhancement
to the Drell-Yan continuum in the IMR, we have to address the
following two questions: (1) can the leading power corrections to
the Drell-Yan cross section be systematically calculated 
in QCD perturbation theory, and (2) is the sign of the nuclear
modification to the Drell-Yan continuum positive.  As we will show
below, answers to these two questions are both positive.

As a direct consequence of the generalized factorization theorem
\cite{QS-fac}, the leading power corrections to the inclusive
Drell-Yan cross section or Drell-Yan continuum can be systematically
calculated in QCD perturbation theory \cite{QS-dy-t4}.  
For the following derivation, we choose a frame in which the hadron
(or nucleus) $A$ is moving along the $+z$-axis, and the $B$ is moving
in $-z$ direction.  We introduce the average nucleon momenta,
$p\equiv P_A/A$ and $p'\equiv P_B/B$.  The $\pm$ components of any
vector $k$ are defined in terms of two lightlike vectors $n$ and
$\bar{n}$ are $k^+=k\cdot n$ and $k^-=k\cdot\bar{n}$.  These lightlike
vectors are chosen such that $\bar{n}\cdot n =1$.  In our choice of
frame, the momentum $p$ is up to a mass term proportional to
$\bar{n}$.

\begin{figure}
\begin{center}
\epsfig{figure=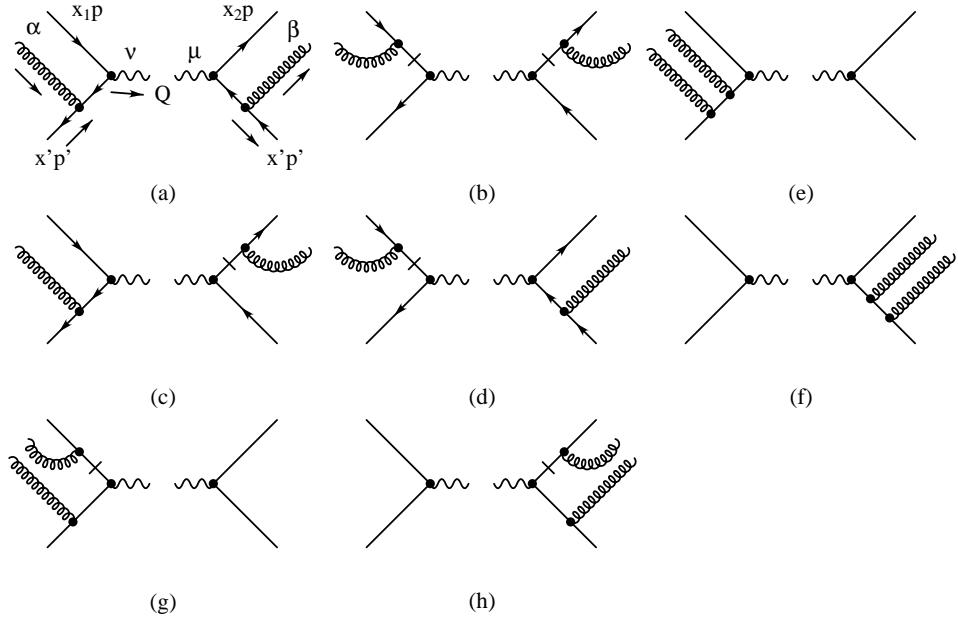,width=5.0in}
\end{center}
\caption{Lowest order Feynman diagrams contribute to the leading power
corrections to Drell-Yan cross section.}
\label{fig2}
\end{figure}

At the leading order in $\alpha_s$, the leading power corrections
to the Drell-Yan cross section are given by the Feynman diagrams
in Fig.~\ref{fig2}, where the quark lines with a short bar are special
quark propagators, which are equal to the normal quark propagators
with the leading twist contributions removed \cite{Qiu-t4}.  According
to the generalized factorization theorem, the leading power
corrections to the Drell-Yan cross section in
hadron-nucleus collisions can be factorized as \cite{QS-dy-t4},
\begin{eqnarray}
W_{\mu\nu}^{(d)}(Q) 
&=& \sum_q
  \int dx' \phi_{\bar{q}/h'}(x')
  \int dx\, dx_1\, dx_2
\nonumber \\
&\times &
  T_{qA}(x,x_1,x_2)\,
  H_{\mu\nu}^{(d)}(x,x_1,x_2;x';Q)\, ,
\label{w-fac-d}
\end{eqnarray}
where $\sum_q$ runs over all quark and antiquark flavors.  In
the light-cone gauge of strong interaction, gluon field strength,
$F^{+\alpha}(y^-)=n^\rho \partial_\rho A^\alpha(y^-)$, and the
hadronic matrix element $T_{qA}(x,x_1,x_2)$ is given by
\cite{BQ-dy-sts}  
\begin{eqnarray}
T_{qA}(x,x_1,x_2)
&=&
\left(\frac{1}{x-x_1}\right) \left(\frac{1}{x-x_2}\right)
\int \frac{dy^-}{2\pi}\, \frac{dy_1^-}{2\pi}\,
      \frac{dy_2^-}{2\pi}
\nonumber \\
&\times &
      {\rm e}^{ix_1p^+y^-}\, {\rm e}^{i(x-x_1)p^+y_1^-}\,
      {\rm e}^{i(x-x_2)p^+y_2^-}
\nonumber \\
&\times &
\langle P_A | \bar{\psi}(0) \frac{\gamma^+}{2}
              F^{+\alpha}(y_2^-)F_{\alpha}^{\ +}(y_1^-)
              \psi(y^-) |P_A\rangle\, .
\label{TqA}
\end{eqnarray}
The partonic part $H_{\mu\nu}^{(d)}$ in
Eq.~(\ref{w-fac-d}) is given by the Feynman diagrams in
Fig.~\ref{fig2}, in which quark (antiquark) lines are contracted by
$\frac{1}{2}\gamma\cdot p$ ($\frac{1}{2}\gamma\cdot p'$) and gluon
indices are contracted with $\frac{1}{2} d_{\alpha\beta}$.  The
transverse polarization tensor $d_{\alpha\beta}=-g_{\alpha\beta}
+\bar{n}_\alpha n_\beta + n_\alpha \bar{n}_\beta$.

Because of EM current conservation, we can calculate the partonic hard
part by contracting the $H_{\mu\nu}^{(d)}$ with either $-g^{\mu\nu}$
or the photon polarization tensor $P^{\mu\nu}(Q)$ in any gauge choice
of EM interaction.  Although the final answer is the same,
contributions from different diagrams to the final answer are
different.  In addition, with an explicit polarization tensor, we can
keep track the contributions to the different polarization states of
the virtual photon.  In the following, we present a result calculated
in the light-cone gauge of EM interaction.  In this gauge, the virtual
photon's polarization tensor is given by \cite{QZ-photon}
\begin{eqnarray}
P^{\mu\nu}(Q) 
&=& -g^{\mu\nu}
    + \frac{Q^\mu n^\nu + n^\mu Q^\nu}{Q\cdot n}
\nonumber \\
&=& 2 P_T^{\mu\nu}(Q) + P_L^{\mu\nu}(Q)\, .
\label{pmn-lc}
\end{eqnarray}
With $Q^\mu = Q^+ \bar{n}^\mu + Q^- n^\mu$ in our frame, we have
\begin{equation}
P_T^{\mu\nu}(Q) = \frac{1}{2}\, d^{\mu\nu}\, ,
\quad \mbox{and} \quad
P_L^{\mu\nu}(Q) = \frac{2Q^-}{Q^+}\, n^\mu\, n^\nu\, .
\label{pmn-t-l}
\end{equation}
Contracting the partonic part $H_{\mu\nu}^{(d)}$ with the polarization
tensors in Eq.~(\ref{pmn-t-l}), we find that only one diagram in
Fig.~\ref{fig2}(a) contributes to the longitudinally polarized virtual
photon state, and two diagrams in Figs.~\ref{fig2}(e) and (f)
contribute to the transversely polarized virtual photon state.  We
obtain 
\begin{eqnarray}
H_{\mu\nu}^{(d)}(Q) P_L^{\mu\nu}(Q)
&=& e_q^2 \left( \frac{1}{18} \right)
    \frac{g^2}{xx'}\;
    (2\pi)^4 S
\nonumber \\
&\times &\,
    \delta(Q^+-xp^+)\,
    \delta(Q^--x'p'^-)\,
    \delta^2(Q_T)\, ,
\label{h-l} \\
H_{\mu\nu}^{(d)}(Q) P_T^{\mu\nu}(Q)
&=& e_q^2 \left( \frac{1}{18} \right)
    \frac{g^2}{xx'}\; (2\pi)^4 S \, 
\nonumber \\
&\times & \,
  x \left[
    \frac{\delta(Q^+-x_2p^+)}{x_2-x_1}
  + \frac{\delta(Q^+-x_1p^+)}{x_1-x_2} \right]
    \delta(Q^--x'p'^-)\,
    \delta^2(Q_T)\, ,
\label{h-t}
\end{eqnarray}
where $g$ is the coupling constant of strong interaction and $(1/18)$
is the overall color factor.  Total partonic part to the inclusive
Drell-Yan cross section is equal to the longitudinal contribution in
Eq.~(\ref{h-l}) plus twice of the transverse contribution in
Eq.~(\ref{h-t}).  If we contract the $H_{\mu\nu}^{(d)}$ with
$-g^{\mu\nu}$, we get the exact same total contributions, except they
come from different diagrams.

As pointed out in Ref.~\cite{QS-hardprobe}, the perturbatively
calculated hard part in the generalized factorization formula is
independent of the structure -- in particular the size -- of the
nuclear target.  We need to find the medium size enhancement due to
multiple scattering from the matrix element in Eq.~(\ref{TqA}).  The
$y^-_i$ integrals in Eq.~(\ref{TqA}) parameterize the distance between
the gluon and quark.  Generally, the integrals over the distance
$y_i^-$ cannot grow with the size of the target because of
oscillations of the exponential factors, unless the integrations of
parton momentum fractions lead to vanishing exponential factors.
This can be achieved if the integrations of parton momentum fractions
are dominated by unpinched poles \cite{QS-hardprobe}.

Combine the partonic parts in Eqs.~(\ref{h-l}) and (\ref{h-t}) and the
matrix element in Eq.~(\ref{TqA}), we find that the $dx_1$ and $dx_2$
integrations are dominated by the region where $x_1\sim x$ and
$x_2\sim x$ from the poles.  That the poles are in the matrix element
$T_{qA}$ and do not have explicit $i\epsilon$ is due to our
choice of light-cone gauge for the strong interaction.  Within the
light-cone gauge, we can either assign the $i\epsilon$ according the
prescription introduced in the Appendix of Ref.~\cite{MQ-recom} or
carefully keep track of the $i\epsilon$ in the partonic part.  For
example, we left the following factor
$$
\left( \frac{x-x_1}{x-x_1+i\epsilon} \right)
\left( \frac{x-x_2}{x-x_2-i\epsilon} \right)
$$
from the right-hand-side of Eq.~(\ref{h-l}).  The numerator cancels
the artificial poles in the matrix element and leaves the poles with
correct $i\epsilon$ to the partonic hard part.  We can also work in a
covariant gauge of strong interaction without worrying about the
artificial poles caused by the choice of light-cone gauge
\cite{BQ-dy-sts,LQS-photon}, and the same answer was derived.

Using the unpinched poles $1/(x-x_1+i\epsilon)$ and
$1/(x-x_2-i\epsilon)$ to carry out the $dx_1$ and $dx_2$ integrations
in Eq.~(\ref{w-fac-d}), and substituting the hadronic tensor
$W_{\mu\nu}^{(d)}$ into Eq.~(\ref{dy-d4q}), we obtain the leading
order medium size enhanced power corrections to the inclusive 
Drell-Yan cross section in hadron-nucleus collisions,
\begin{equation}
\frac{d\sigma_{Ah'}^{(d)}}{dQ^2}
\approx 
  \left( \frac{4\pi \alpha_{em}^2}{9Q^2} \right)
  \sum_{q}\, e_q^2\,
  \int dy\, 
\left(\frac{4\pi^2}{3}\,\frac{\alpha_s}{Q^2}\right)
\left[2\; T_{qF}(x)-x\frac{\partial}{\partial x}T_{qF}(x)\right]
\phi_{\bar{q}/h'}(x')
\label{dy-dq2-d-lo}
\end{equation}
with $x$ and $x'$ given in Eq.~(\ref{x-xp}).  The term proportional to
$T_{qF}$ (its derivative) contributes 
to the longitudinally (transversely) polarized virtual photon state.
The twist-4 quark-gluon correlation function $T_{qF}(x)$ is defined as
\cite{LQS-photon,Guo-dy-qt2}
\begin{eqnarray}
T_{qF}(x)
&=&
 \int \frac{dy^-}{2\pi}\, {\rm e}^{ixp^+y^-}
 \int \frac{dy_1^-dy_2^-}{2\pi}\,
      \theta(-y_2^-)\theta(y^--y_1^-)
\nonumber \\
&\times &
\langle P_A | \bar{\psi}(0) \frac{\gamma^+}{2}
              F^{+\alpha}(y_2^-)F_{\alpha}^{\ +}(y_1^-)
              \psi(y^-) |P_A\rangle .
\label{TqF}
\end{eqnarray}
In deriving Eq.~(\ref{dy-dq2-d-lo}), we used the following
approximation in the limit $x_1\rightarrow x$ and $x_2\rightarrow x$,
\begin{eqnarray}
&\ &
\left[ \frac{\delta(x_2-Q^+/p^+)}{x_2-x_1}
      +\frac{\delta(x_1-Q^+/p^+)}{x_1-x_2} \right]
   \left(\frac{1}{x-x_1+i\epsilon}\,
         \frac{1}{x-x_2-i\epsilon}\right)
\nonumber \\
&\approx &\,
   \delta'(x-Q^+/p^+)
   \left(\frac{1}{x-x_1+i\epsilon}\, 
         \frac{1}{x-x_2-i\epsilon}\right)
   + ...
\label{d-delta}
\end{eqnarray}
where ``...'' represents the terms that have less than two poles.

At the leading order in $\alpha_s/Q^2$, the medium enhanced power
corrections to the inclusive Drell-Yan cross section in
nucleus-nucleus collisions are given by
\begin{equation}
\frac{d\sigma_{AB}^{(d)}}{dQ^2}
\approx
  B \frac{d\sigma_{AN}^{(d)}}{dQ^2}
+ A \frac{d\sigma_{NB}^{(d)}}{dQ^2}\, ,
\label{dy-dq2-d-ab}
\end{equation}
with the nucleon-nucleus contributions given in
Eq.~(\ref{dy-dq2-d-lo}).

In order to know the sign and magnitude of the medium enhanced power
corrections, we need to know both the correlation function,
$T_{qF}(x)$, and its derivative.  Although the correlation function is 
nonperturbative, the generalized factorization theorem requires it to
be a universal function.  Therefore, we can find the information of
$T_{qF}(x)$ from other observables.

In Ref.~\cite{Guo-dy-qt2}, Drell-Yan lepton-pair's transverse momentum
broadening was calculated in terms of parton multiple scattering in
perturbative QCD.  It was found that the broadening is proportional to
the same quark-gluon correlation $T_{qF}(x)$.  Using data on Drell-Yan
transverse momentum broadening in hadron-nucleus collisions, and a
model for $T_{qF}(x)$ from Ref.~\cite{LQS-photon},
\begin{equation}
T_{qF}(x) \approx \lambda^2\, A^{4/3}\, \phi_{q/N}(x)\, ,
\label{TqF-model}
\end{equation}
it was found \cite{Guo-dy-qt2} that $\lambda^2\approx 0.01$~GeV$^2$.
The same model was also used to evaluate 2-jet momentum imbalance in
photon-nucleus collision \cite{LQS-photon}, and a larger value of
$\lambda^2$ ($\sim 0.05$~GeV$^2$) was extracted from Fermilab data
\cite{E683}.  Differences between these two observables were discussed
in Ref.~\cite{QS-hardprobe}. 

Taking the same model for the quark-gluon correlation function in
Eq.~(\ref{TqF-model}), and substituting Eq.~(\ref{dy-dq2-d-ab}) into
Eq.~(\ref{rat-d-s}), we obtain
\begin{eqnarray}
R_{AB}(Q)
&\approx &
\frac{4\pi^2\alpha_s}{3Q^2}\, \lambda^2
\nonumber \\
&\times &
\frac{\sum_q e_q^2 \int dy\, 
           \left\{ A^{1/3}
           \left[2\, \phi_{q/N}(x)
                -x\frac{\partial}{\partial x}\phi_{q/N}(x)\right]
                \phi_{\bar{q}/N}(x') 
              + B^{1/3}[x\leftrightarrow x']\right\}}
     {\sum_q e_q^2 \int dy\, \phi_{q/N}(x)\,\phi_{\bar{q}/N}(x')}\, ,
\label{R-AB-model}
\end{eqnarray}
where $\sum_q$ runs over all quark and antiquark flavors, and $x$ and
$x'$ are given in Eq.~(\ref{x-xp}).

\begin{figure}
\begin{center}
\epsfig{figure=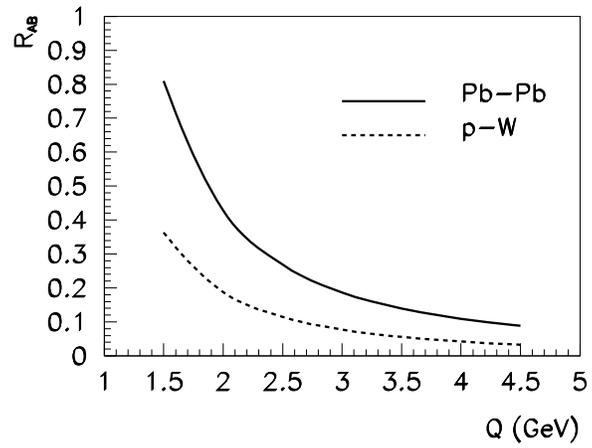,width=3.0in}
\end{center}
\caption{The ratio $R_{AB}(Q)$ in Eq.~(\protect\ref{rat-d-s}) as a
function of dilepton mass $Q$ in the Pb-Pb (solid) and $p-W$ (dashed)
collisions.}  
\label{fig3}
\end{figure}

To be consistent with our leading order calculation in $\alpha_s$, we
use leading order parton distributions CTEQ5L for the 
numerical evaluation of the ratio $R_{AB}(Q)$ \cite{CTEQ5-pdf}.
Although the value of the $\lambda^2$ is not well fixed, we will use
$\lambda^2=0.01$~GeV$^2$ to calculate the $R_{AB}(Q)$, because it was 
extracted from a similar process -- Drell-Yan transverse momentum
broadening.  The uncertainty in the value of $\lambda^2$
corresponds to a shift in overall normalization of the power
corrections.  we present the ratio $R_{AB}(Q)$ as a function of
dilepton's invariant mass $Q$ in Fig.~\ref{fig3}.  The solid line is
for Pb-Pb collision at $\sqrt{S}=17.8$~GeV/nucleon with c.m. rapidity
coverage $0<y_{cm}<1$, while the dashed line is for p-W collision at
$\sqrt{S}=30$~GeV/nucleon with c.m. rapidity coverage
$-0.52<y_{cm}<0.48$.  

From the dashed line in Fig.~\ref{fig3}, the nuclear enhancement to
the Drell-Yan continuum in the $p-W$ collision can be as large as 20\%
at $Q\sim 2$~GeV.  This is not in any contradiction with the data on
the opposite-sign dilepton production from NA38 and NA50
collaborations, which show a good agreement with the expectation from
all sources \cite{NA38NA50-dimuon}.  As shown in
Ref.~\cite{NA38NA50-dimuon}, a majority of lepton pairs in this IMR in 
hadron-nucleus collisions at $\sqrt{S}=30$~GeV come from open charm
$D\bar{D}$ decay.  Actually, a simultaneous fit of the 4 set $p-A$
data shows that the ratio of opposite-sign dileptons from
the open charm decay and those directly from the Drell-Yan process is
as large as $4.2 \pm 0.9$ at $\sqrt{S}=30$~GeV \cite{NA38NA50-dimuon}.
Therefore, a 20\% increase of lepton pairs from the nuclear
enhancement of the Drell-Yan continuum corresponds to less than 5\%
uncertainty in lepton pairs from $D\bar D$ decay, which is within the
error bar of the fitting. 

On the other hand, as shown in Fig.~9 of Ref.~\cite{NA38NA50-dimuon},
the expected opposite-sign lepton pairs from the $D\bar{D}$ decay are
less than the pairs from the Drell-Yan continuum in the Pb-Pb
collision.  This is because total production rate of open charm
$D\bar{D}$ pairs are much more suppressed than that of the Drell-Yan
continuum when the collision energy decreases from $\sqrt{S}=30$~GeV
for $p-A$ to $\sqrt{S}=17.8$~GeV for the Pb-Pb collision.  With the
relative dominance of the Drell-Yan continuum in this IMR, a 50\%
enhancement of the Drell-Yan continuum at $Q\sim 2$~GeV from the solid
line in Fig.~\ref{fig3} is a very significant change for the expected
dilepton sources.  Because of the mass threshold for producing open
charm $D\bar{D}$ pairs, we expect the power corrections to the
dilepton source from open charm decay to be much smaller.  

We also emphasize that when the dilepton's invariant mass $Q$ is
much less than 2~GeV, the leading order power corrections, as shown in
Fig.~\ref{fig3}, become so large that higher order power corrections
have to be considered for a reliable QCD prediction.  In addition,
there should be high order corrections in $\alpha_s$ to the
leading power contributions as well as the power corrections.  Since
we are interested in the relative size of these two contributions, we
expect the ratio $R_{AB}(Q)$ to be relatively stable when we add high
order corrections in $\alpha_s$.  Finally, we point out that 
sign of the power corrections is not necessary to be positive as what
we find here.  Actually, as shown in Ref.~\cite{GQZ-dis}, the power
corrections to the longitudinal and transverse structure functions in
deeply inelastic scattering carry an opposite sign.

In conclusion, we calculated the nuclear size enhanced
power corrections to the inclusive Drell-Yan cross section in the IMR
in hadron-nucleus and nucleus-nucleus collisions.  We find that the
medium size enhanced power corrections significantly enhance the
dilepton production in the IMR, in particular, in the Pb-Pb collision.
Therefore, in order to understand the measured nuclear enhancement of
lepton pairs in the IMR at CERN SPS, a new fit for the expected
sources including the power corrections to the Drell-Yan continuum
will be very important. 

\vskip 0.2in

This work was supported in part by the U.S. Department of Energy under
Grant Nos. DE-FG02-86ER40251 and DE-FG02-87ER40731.



\end{document}